\documentclass[useAMS,usenatbib,usegraphicx]{mn2e}

\title[FOS Ly$\alpha $ Forest Absorption]{Continuous statistics of the 
      \lyaf\ at $0 < z < 1.6$: the mean flux,  flux distribution,  
      and autocorrelation from HST FOS spectra.}
\author[D. Kirkman \etal]{David Kirkman$^{1}$\thanks{E-mail: dkirkman@ucsd.edu}
  David Tytler$^{1}$, 
  Dan Lubin$^{1}$, 
  and Jane Charlton$^{2}$
  \\
  $^{1}$ Center for Astrophysics and Space Sciences, 
  University of California San Diego,
  La Jolla, CA, 92093-0424
  \\
  $^{2}$ 525 Davey Lab, Dept. of Astronomy and Astrophysics, Penn State, 
  University Park, PA, 16802
}

\usepackage{color}
\usepackage{graphicx}
 
\newcommand{\lya}{\mbox{Ly$\alpha$}}
\newcommand{\lyb}{\mbox{Ly$\beta$}}

\newcommand{\kms}{\mbox{km s$^{-1}$}}
\newcommand{\cmm}{\mbox{cm$^{-2}$}}

\newcommand{\ob}{\mbox{$\Omega_b$}}

\newcommand{\zem}{\mbox{$z_{\rm em}$}}

\newcommand{\nhi}{\mbox{N$_{\rm H I}$}}
\newcommand{\lnhi}{\mbox{log \nhi}}

\newcommand{\etal}{{\it et al.}}
\newcommand{\lyaf} {\lya\ forest}

\newif\ifdraftmodep
\draftmodepfalse

\newif\ifapjp
\apjpfalse

\begin{document}

\date{\today}
 
\maketitle

\begin{abstract}

  We measure the amount of absorption in the \lyaf\ at $0 < z < 1.6$
  in Hubble Space Telescope Faint Object Spectrograph spectra of 74
  QSOs.  Starting with a 334 QSO sample compiled by Bechtold
  \etal\ 2002, we selected 74 QSOs that have the highest signal to
  noise and complete coverage of rest frame wavelengths 1070
  --1170~\AA .  We measure the absorption from the flux in each pixel
  in units of the unabsorbed continuum level. We mask out regions of
  spectra that contain metal lines, or strong \lya\ lines that are
  accompanied by other Lyman series line or metals at the same
  redshift, leaving \lya\ absorption from the low density
  intergalactic medium.  At $0 < z < 1.6$ we find that 79\% of the
  absorption is from the low density intergalactic medium, 12\% from
  metals and 9\% from the strong H~I lines, nearly identical to the
  percentages (78, 15 and 7) that we measured independently at $z=2$
  from spectra taken with the Kast spectrograph on the Lick 3-m. At
  $z=1$ the low density intergalactic medium absorbs $0.037 \pm 0.004$
  of the flux. The error includes some but not all of the uncertainty
  in the continuum level. The remaining part gives relative errors of
  approximately 0.21 when we report the mean absorption in eight
  independent redshift intervals, and 0.047 when we average over all
  redshifts.  We find 1.46 times more absorption from the low density
  intergalactic medium than comes from \lya\ lines that Bechtold
  \etal\ 2002 listed in the same spectra.  The amount of absorption
  increases with $z$ and can be fit by a power law in $(1+z)$ with
  index 1.01. This corresponds to no change in the number of lines, of
  fixed rest frame equivalent widths, per unit redshift, consistent
  with the Janknecht \etal\ 2006 results on the distribution of lines.
  When we include similar measurements from higher redshifts, we need
  more degrees of freedom to fit the amount of absorption at $0 < z <
  3.2$. A power law with a break in slope, changing from index 1.5 at
  low $z$ to 3.0 above $z \sim 1.1$ is a better but only marginally
  acceptable fit.  We also calculate two other continuous statistics,
  the flux probability distribution function and the flux
  autocorrelation function that is non-zero out to $v \sim 500$ \kms\
  at $0.5 < z < 1.5$.

\end{abstract}

\begin{keywords}
quasars: absorption lines -- cosmology: observations -- intergalactic medium.
\end{keywords}

\section{Introduction}

The \lyaf\ seen in the spectra of distant QSOs is a key probe of the
intergalactic medium (IGM) at redshifts $z < 6$.  The \lyaf\ is
produced by neutral Hydrogen and so probes both the distribution
of matter and the ionization state of the IGM.  We have been working
on extracting the properties of the IGM from detailed comparisons of
\lyaf\ observations with grids of numerical cosmological simulations
\citep{tytler04b, jena05a, kirkman05a}. 

The simplest measurement to make of the \lyaf\ is the average
absorption.  This is sometimes reported as a mean flux $\overline{F}$
or as an effective opacity $\tau_{\rm eff}$, but in this paper we
follow the convention of \citet{oke82} and report DA
$=1-\overline{F}$, where DA stands for the flux Decrement in the Lyman
Alpha region of a QSO spectrum. DA and its redshift evolution have
long been known to place important constraints on the properties of
the IGM \citep{jenkins91}. Numerous authors have noted that
measurements of DA can constrain key cosmological and astrophysical
parameters such as the baryon density \ob, $\sigma_8$, the temperature
density relationship of the IGM, and the intensity of the UV
background $\Gamma_{912}$ \citep{hernquist96, rauch97, rauch98,
  mcdonald01b, tytler04b}.  Recent work at high redshift has focused
on using the evolution of DA as a probe of the ionization state of the
IGM \citep{songaila04a, bernardi03}.  We \citep{tytler04b, jena05a,
  kirkman05a} used the combination of DA and the distribution of line
widths to show which combinations of cosmological and astrophysical
parameters give numerical simulations of the \lyaf\ that best match
spectra.

DA or the mean flux is an example of a continuous statistic that is
defined at all wavelengths in a spectrum. A complementary set of
discrete statistics are also widely used. These begin with a list of
absorption lines in a spectrum, each of which has fitted properties,
such as equivalent width W, and central wavelength. One such line 
counting statistic is the number of lines with W exceeding some minimum 
as a function of redshift.

Most prior work at low redshifts used line counting \citep{ bahcall93,
  bahcall96, weymann98, jannuzi98, dobrzycki02, janknecht06a}.  The
decrease in the number of lines near to QSOs has been used to estimate
the intensity of the ultraviolet background (UVB) \citep{ kulkarni93,
  shull99, scott00b, scott02}.  At the lowest redshifts, QSOs and
Seyfert galaxies are bright enough to allow high resolution spectra
that also give line widths and hence column densities (Shull \etal\
1999; Penton \etal\ 2000a, 2000b, 2004).  \nocite{shull99b, penton00a,
  penton00b, penton04a}

A major early result from HST QSO spectra was
that the number of lines seen at $z < 1.7$ was significantly more than
expected from an extrapolation of the trend measured at higher
redshifts.  \citet{dave99} showed that in the context of numerical IGM
simulations the evolution of the line counts was driven by a
combination of cosmological expansion (which tends to increase
ionization) and the fading of the UV background (which decreases H~I
ionization).  The number of lines counted as a function of redshift is
often fit with a broken power law with a shallower slope below $z \sim
1.7$. This is approximately the redshift where the ionization changes
caused by the fading UV background become comparable with the
ionization changes caused by the general expansion. At lower redshifts
the drop in the UVB tends to cancel the expansion, leading to
decreased evolution at lower redshifts.
In more detail, the redshift of the break in slope is poorly known, and can be
much lower than $z=1.7$. Recently \citet{janknecht06a} 
claimed the break is at $z \sim 0.7$.

The flux probability distribution function (PDF) gives the probability
that a pixel will have a given flux value on a scale from zero to one at 
the continuum flux. This was first measured in the \lyaf\ by \citet{jenkins91}.
More recently, the flux PDF at $z > 2$ has been derived from Keck +
HIRES spectra \citep{mcdonald00a, becker06a} and from a sample of SDSS
quasars spectra \citep{desjacques06a}.  \citet{becker06a} found that a
simple description of the \lyaf\ using a log-normal density distribution
\citep{bi93, bi97a} accurately reproduced the \lyaf\ flux PDF between
$1.7 < z < 5.8$. But \citet{desjacques06a} also found that the same
log-normal model failed to describe a large sample of SDSS spectra over a
similar redshift range, unless they assumed surprisingly
large 20\% errors in the continuum levels used for the SDSS spectra. 

In this paper, we use Hubble Space Telescope (HST) Faint Object Spectrograph 
(FOS) spectra to measure DA, the flux PDF and the
flux autocorrelation function, a continuous version of the two point
correlation function that is often applied to samples of absorption
lines. We also combine these DA measurements with our earlier measurements 
at higher redshifts up to $z = 3.2$.

\section{HST QSO Spectra}

We measure the properties of the \lyaf\ at rest frame wavelengths
1070 -- 1170~\AA\ in the spectra of bright QSOs. At low redshifts we must use UV
spectra of QSOs, nearly all of which are from HST, since IUE spectra
are largely superseded by HST spectra of the same QSOs. We considered HST
spectra from FOS, GHRS and STIS, and chose the FOS spectra for this project. 
The moderate resolution GHRS
spectra with spectral resolution R=21,000 are of little use because
they have a small wavelength coverage.  There are about a dozen QSOs
with suitable high resolution STIS spectra 
\citep{milutinovic06a, janknecht06a} with
R = 30,000. We will examine several of these to check 
our FOS results, but otherwise they are too few. We
know from \cite{tytler04a} that we need many tens of spectra to
average over the huge variations in the amount of absorption from QSO
to QSO, and only FOS has observed this many QSOs.
We did not consider the low resolution STIS spectra.


We use the atlas of 334 HST FOS spectra published by
\citet[B02]{bechtold02}, who collected all of the 
spectra of QSOs obtained with the FOS G130H, G190H, or G270H gratings
and reduced with the STSDAS package.
The spectra have a pixels of size 
0.25~\AA\ (G130H), 
0.37~\AA\ (G190H) and 
0.52~\AA\ (G270H), and 4 pixels per diode.
These gratings together provide overlapping wavelength coverage from
below 1200 to 3200~\AA\ at a spectral resolution of approximately
R=1300 (FWHM 230~\kms ) for the usual aperture width and guiding
jitter. 

For analysis, we selected spectra from the B02 archive that had an
average signal-to-noise ratio SNR $> 20$ per pixel between 1070 --
1170~\AA, and complete coverage of the \lyaf\ between the \lya\ and
the \lyb\ emission lines of the QSO. These two criteria leave 71 QSOs
that we list in Table \ref{fosspectab}. We deliberately added three
more QSOs with lower SNR (3C298, Q1026-0045B, Q1258+285) to give more
spectra at the highest redshifts and 74 in total.  These are amongst
the brightest QSOs in the sky at low redshifts, and we are not aware
of any factors in their selection that would bias a DA measurement. We
restrict the sample to QSOs with high SNR and complete coverage of the
\lyaf\ because we know that the continua that we fit have
significantly higher errors otherwise.

Most of the QSOs listed in Table \ref{fosspectab} have spectra from
multiple FOS gratings.  To combine the data sets from different
gratings, we re-sampled the spectrum from each grating onto a common
wavelength scale, and then combined the separate data sets pixel by
pixel.  For the pixels in the overlap region between gratings, we took
the SNR weighted mean of the two observations.  The B02 spectra are
well calibrated, and there are no apparent artifacts in the spectrum
overlap region between adjacent gratings.

\begin{table}
\caption{The 74 QSOs with FOS spectra used to measure the \lyaf\ absorption.  
         QSOs for which we have also analyzed STIS spectra are indicated 
         with a $^1$ after their name.  $z_{\rm forest}$ is the mean redshift
         of the \lyaf\ region analyzed.
}
\label{fosspectab}
\begin{tabular}{l cc cc c}
\hline
Name & RA & Dec & $z_{\rm em}$  &  $z_{\rm forest}$  & LyaF\cr
     & (J2000) & (J2000) & & & SNR \cr
\hline
MARK132 & 10:01:29.8 & +54:54:39 & 1.750 & 1.534  &    25 \cr
PG1115+080A1 & 11:18:16.9 & +07:45:59 & 1.722 & 1.508  &    24 \cr
Q1026$-$0045$-$B & 10:28:37.0 & $-$01:00:28 & 1.530 & 1.331  &    13 \cr
PKS0743$-$67 & 07:43:31.7 & $-$67:26:25 & 1.511 & 1.313  &    22 \cr
PG0117+213$^1$ & 01:20:17.3 & +21:33:46 & 1.493 & 1.297  &    39 \cr
PG1630+377 & 16:32:01.1 & +37:37:49 & 1.478 & 1.283  &    37 \cr
HS$-$1216+5032A & 12:18:41.1 & +50:15:36 & 1.450 & 1.257  &    22 \cr
PKS0232$-$04 & 02:35:07.3 & $-$04:02:06 & 1.450 & 1.257  &    34 \cr
3C298 & 14:19:08.2 & +06:28:34 & 1.436 & 1.244  &    18 \cr
0957+561B & 10:01:20.9 & +55:53:49 & 1.414 & 1.224  &    29 \cr
1E0302$-$223 & 03:04:49.8 & $-$22:11:52 & 1.400 & 1.211  &    28 \cr
QSO$-$1258+285 & 13:01:00.9 & +28:19:44 & 1.355 & 1.170  &    16 \cr
0454+0356 & 04:56:47.2 & +04:00:53 & 1.345 & 1.160  &    29 \cr
PG1634+706$^1$ & 16:34:29.1 & +70:31:32 & 1.337 & 1.153  &   115 \cr
PG1522+101 & 15:24:24.5 & +09:58:29 & 1.324 & 1.141  &    47 \cr
Q1435$-$0134 & 14:37:48.3 & $-$01:47:11 & 1.310 & 1.128  &    44 \cr
PG1008+133 & 10:11:10.8 & +13:04:11 & 1.287 & 1.107  &    31 \cr
PG1241+176$^1$ & 12:44:10.8 & +17:21:03 & 1.273 & 1.094  &    30 \cr
4C06.41 & 10:41:17.2 & +06:10:17 & 1.270 & 1.091  &    30 \cr
PKS$-$1327$-$206 & 13:30:07.7 & $-$20:56:17 & 1.169 & 0.998  &    28 \cr
PG1718+481$^1$ & 17:19:38.3 & +48:04:12 & 1.084 & 0.920  &    83 \cr
PKS1229$-$02 & 12:32:00.0 & $-$02:24:05 & 1.045 & 0.884  &    22 \cr
PG1248+401$^1$ & 12:50:48.3 & +39:51:40 & 1.030 & 0.870  &    30 \cr
TON153 & 13:19:56.3 & +27:28:08 & 1.022 & 0.863  &    46 \cr
PKS2145+06 & 21:48:05.5 & +06:57:38 & 0.990 & 0.833  &    28 \cr
1148+5454 & 11:51:20.4 & +54:37:32 & 0.969 & 0.814  &    29 \cr
TON157 & 13:23:20.5 & +29:10:07 & 0.960 & 0.806  &    25 \cr
Q0107$-$025A & 01:10:13.1 & $-$02:19:54 & 0.948 & 0.795  &    25 \cr
PG1407+265 & 14:09:23.9 & +26:18:21 & 0.944 & 0.791  &    50 \cr
Q0107$-$025B & 01:10:16.3 & $-$02:18:53 & 0.942 & 0.789  &    29 \cr
PKS2340$-$03 & 23:42:56.6 & $-$03:22:26 & 0.896 & 0.747  &    42 \cr
PKS1252+11 & 12:54:38.2 & +11:41:06 & 0.871 & 0.724  &    23 \cr
3C454$-$3 & 22:53:57.8 & +16:08:53 & 0.859 & 0.713  &    29 \cr
CSO176 & 12:52:25.0 & +29:13:21 & 0.820 & 0.677  &    27 \cr
PKS1424$-$11 & 14:27:38.1 & $-$12:03:49 & 0.806 & 0.664  &    26 \cr
3C110 & 04:17:16.7 & $-$05:53:46 & 0.781 & 0.641  &    36 \cr
CSO179 & 12:53:17.5 & +31:05:50 & 0.780 & 0.640  &    34 \cr
CT336 & 01:04:41.0 & $-$26:57:08 & 0.780 & 0.640  &    24 \cr
0959+68W1 & 10:03:06.8 & +68:13:17 & 0.773 & 0.633  &    26 \cr
PG1538+477 & 15:39:34.8 & +47:35:30 & 0.770 & 0.631  &    37 \cr
OS562 & 16:38:13.5 & +57:20:23 & 0.745 & 0.608  &    20 \cr
PKS1354+19 & 13:57:04.5 & +19:19:06 & 0.719 & 0.584  &    23 \cr
PKS2352$-$34 & 23:55:25.6 & $-$33:57:55 & 0.706 & 0.572  &    27 \cr
B20923+39 & 09:27:03.0 & +39:02:20 & 0.699 & 0.565  &    29 \cr
PKS2344+09 & 23:46:36.9 & +09:30:46 & 0.672 & 0.540  &    31 \cr
3C57 & 02:01:57.2 & $-$11:32:34 & 0.669 & 0.538  &    52 \cr
PKS0637$-$75 & 06:35:46.7 & $-$75:16:16 & 0.656 & 0.526  &    20 \cr
3C263 & 11:39:57.1 & +65:47:49 & 0.652 & 0.522  &    32 \cr
MC1104+167 & 11:07:15.0 & +16:28:02 & 0.634 & 0.505  &    37 \cr
PKS2243$-$123 & 22:46:18.2 & $-$12:06:51 & 0.630 & 0.502  &    39 \cr
PKS0044+030 & 00:47:05.9 & +03:19:55 & 0.624 & 0.496  &    21 \cr
3C95 & 03:51:28.6 & $-$14:29:09 & 0.616 & 0.489  &    51 \cr
4C41.21 & 10:10:27.5 & +41:32:39 & 0.611 & 0.484  &    36 \cr
PKS0405$-$12 & 04:07:48.4 & $-$12:11:36 & 0.574 & 0.450  &    66 \cr
PKS0454$-$22 & 04:56:08.9 & $-$21:59:09 & 0.534 & 0.413  &    37 \cr
PKS0003+15 & 00:05:59.2 & +16:09:49 & 0.450 & 0.336  &    42 \cr
HS0624+6907 & 06:30:02.7 & +69:05:03 & 0.374 & 0.266  &    61 \cr
PG1216+069 & 12:19:20.9 & +06:38:39 & 0.334 & 0.229  &    42 \cr
TON28 & 10:04:02.6 & +28:55:35 & 0.329 & 0.224  &    55 \cr
PKS2251+11 & 22:54:10.4 & +11:36:38 & 0.323 & 0.219  &    30 \cr
3C249.1 & 11:04:13.9 & +76:58:58 & 0.311 & 0.208  &    50 \cr

\end{tabular}
\end{table}

\begin{table}
\caption{Table \ref{fosspectab} (continued)}
\begin{tabular}{l cc cc c}
\hline
Name & RA & Dec & $z_{\rm em}$  &  $z_{\rm forest}$  & LyaF \cr
     & (J2000) & (J2000) & & & SNR \cr
\hline
4C73.18 & 19:27:48.5 & +73:58:01 & 0.302 & 0.200  &    36 \cr
B22201+31A & 22:03:15.0 & +31:45:37 & 0.297 & 0.195  &    75 \cr
E1821+643 & 18:21:57.2 & +64:20:36 & 0.297 & 0.195  &    95 \cr
PG1444+407 & 14:46:45.9 & +40:35:07 & 0.267 & 0.167  &    43 \cr
PG0953+414 & 09:56:52.4 & +41:15:23 & 0.239 & 0.141  &    46 \cr
PG1427+480 & 14:29:43.1 & +47:47:26 & 0.221 & 0.125  &    39 \cr
PG0947+396 & 09:50:48.4 & +39:26:51 & 0.206 & 0.111  &    38 \cr
PG1322+659 & 13:23:49.5 & +65:41:48 & 0.168 & 0.076  &    45 \cr
1402+2609 & 14:05:16.2 & +25:55:33 & 0.164 & 0.072  &    52 \cr
PG1226+023 & 12:29:06.7 & +02:03:08 & 0.158 & 0.067  &   291 \cr
PG1115+407 & 11:18:30.3 & +40:25:55 & 0.154 & 0.063  &    40 \cr
0026+129 & 00:29:13.8 & +13:16:03 & 0.142 & 0.052  &    59 \cr

\end{tabular}
\end{table}

\section{Preparation of the Spectra}

We fit continua to the FOS spectra manually, using B-splines and the
software described in \citet{kirkman05a}.  Two of us independently (DK
and DT) fit continua to each spectrum. We examined the fits using
plots with various aspect ratios, wavelength scales, and ordering. We
discussed the fits and the differences and then we independently made
slight adjustments. We repeated this process four times until we
agreed, either in the continuum level, or that there was more than one
reasonable interpretation of a spectral feature.  Typically this comes
down to a decision on whether a given region contains a weak emission
line, correlated photon noise, or a group of weak absorption lines.
We found that both the discussions and iterations were essential.

We developed a procedure to make the continua levels consistent among
the QSOs.  We plot the whole spectrum and an enlargement of only the
region of interest.  We fit the continua to spectra displayed with a
bin size of 1~\AA\ for all HST gratings because this larger pixel
reduces the clutter of having 4 pixels per resolution element, and
helps us see the continuum level.  We also plot the residuals from the
continua in units of the error in each pixel, and we work in both
observed and rest wavelengths.  We check, for example, that the DA and
the residuals from the continuum fit do not depend on rest
wavelength. We return to our continuum fits and the likely errors in
them in \S \ref{dafosdata}, \ref{discuss}.

Because we are primarily interested in H I absorption from the low
density IGM, we identified regions of each spectrum suspected of
having either metal or high column density H I absorption.  We first
flagged wavelengths at $\pm 3$~\AA\ of $z=0$ Galactic metal absorption
in every spectrum, whether or not we could see such absorption in a
particular spectrum. We then flagged all pixels within $\pm 3$~\AA\ of
each metal line identified by B02. We analyze the spectra with
different regions included or excluded. If we exclude a pixel from the
sum used to measure a DA value, we call this a masked region.

\begin{figure}
  \includegraphics[width=84mm]{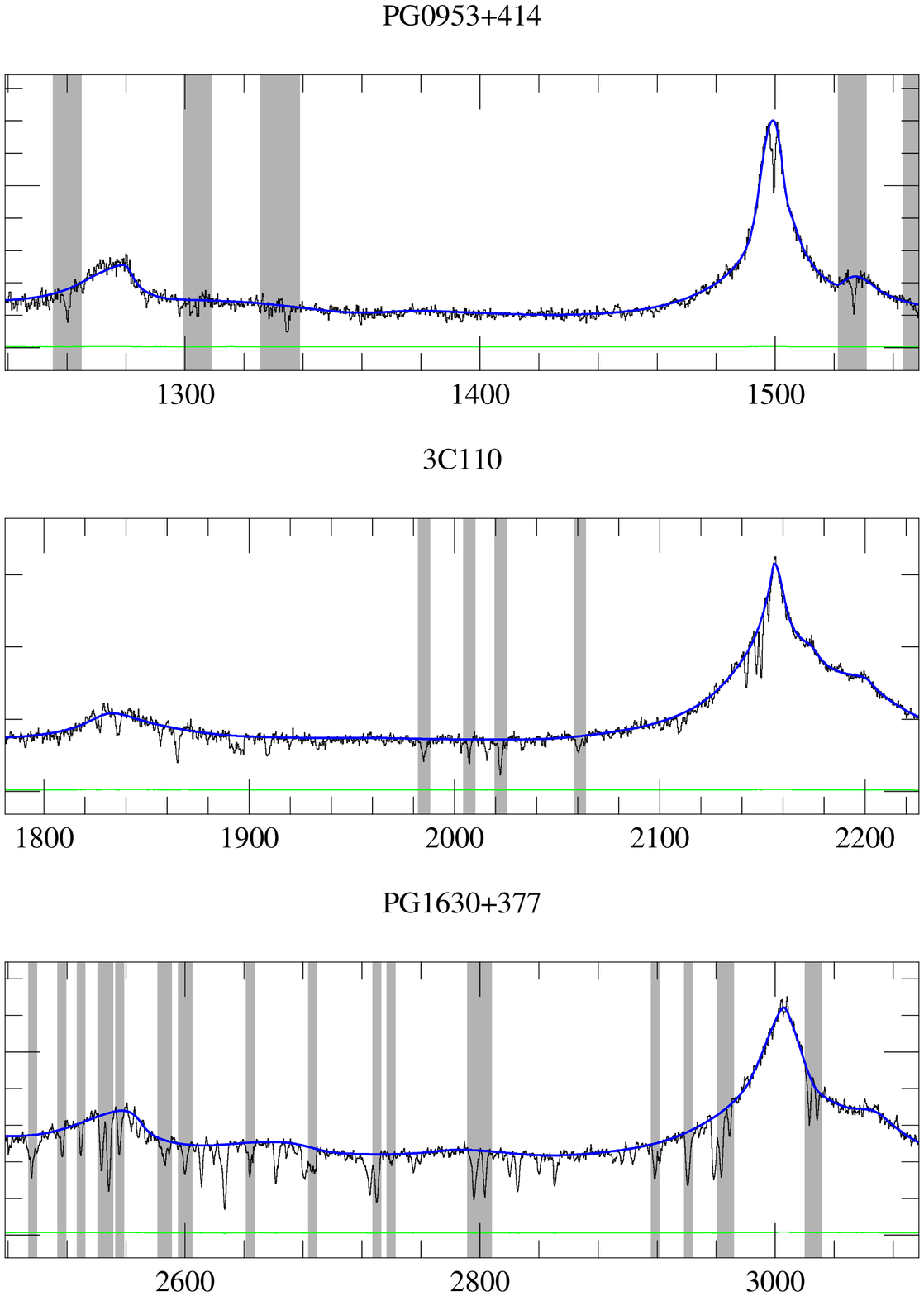}
  \caption{\label{figfosspec} Examples of FOS spectra with
    continua. The wavelengths are observed frame \AA ngstroms, and the
    vertical scale is linear flux with zero at the lowest long
    mark. The thin nearly flat line immediately above the zero level
    is the 1$\sigma $ error.  These three QSOs
    have a range of redshifts and all have typical SNR.  The shaded
    regions are pixels that we flag because they are within 3~\AA\ of
    Galactic and other metal lines and strong H~I lines.  Some regions
    appear wider because there are several masked features near to 
    each other (as in muli-component metal line systems).}
\end{figure}

We define strong H~I lines as any that B02 had identified as
H~I. These are typically absorption systems with higher H~I columns
which show either metal lines in low resolution FOS spectra or Lyman
series lines. Fewer than half of these strong H~I lines will be in
Lyman limit systems (LLS).  This procedure will not remove a
statistically complete sample of high column density H~I lines. It is
a compromise to make use of the limited information in the FOS
spectra. We mask regions within $\pm 3$~\AA\ of each identified metal
line and the strong H~I lines.  In some cases, such as for damped
\lya\ lines, we mask a wider region of the spectrum to cover the
absorption feature.  In Fig. \ref{figfosspec} we show examples of
spectra with continua and flagged regions.

\section{DA in the FOS Spectra}
\label{dafosdata}

We define ${\rm DA} =\left< 1-f_i/c_i \right>$ where $f_i$ is the flux
in pixel $i$ and $c_i$ is the continuum level in the same pixel.  We
restrict our analysis to the rest frame interval 1070 -- 1170~\AA ,
which we divide into three intervals 1070 --
1103\AA, 1103 -- 1137\AA, and 1137 -- 1170\AA.  This gives 3
measurements of DA per QSO. This is a change from our choice at higher
redshifts where we averaged the DA over intervals of redshift path
0.1, or 121.56~\AA\ in the observed frame, or $121.56/(1+z)$ in the
rest frame. At redshift $z=1.7$ these intervals are very similar to
those we now use for HST spectra with fixed rest frame.  We discard
wavelength intervals with $<70$\% of pixels remaining after 
the above masking procedure.

\begin{figure}
   \includegraphics[width=84mm]{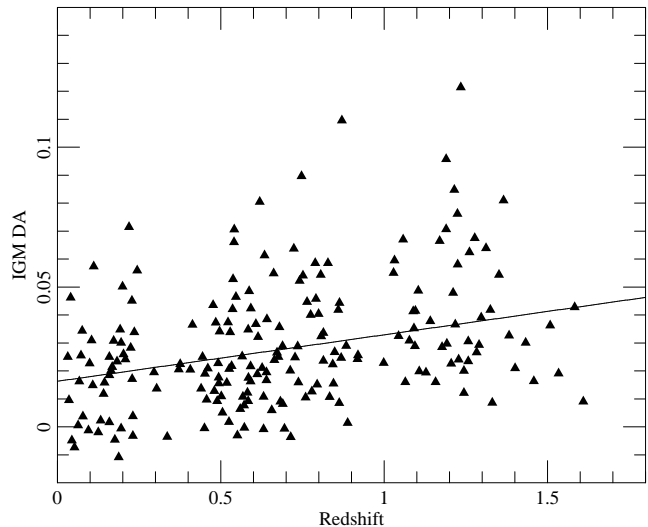}
   \caption{\label{fosonlyda} DA of the low density IGM vs. redshift.
     Metal lines and strong \lya\ lines do not contribute because we 
     masked those wavelengths. Each point
     shows the mean DA in 33.3~\AA\ in the rest frame, with up to 3
     points per QSO.  The line is the best fit power law ${\rm DA}(z)
     =0.016(1+z)^{1.01}$.}
\end{figure}

In Figure \ref{fosonlyda} we show the DA from the low density IGM
alone as a function of redshift. The points shown do not include
pixels masked because they are flagged as metals or strong H~I lines.
All 74 QSOs contribute to this figure, but for many QSOs one or
occasionally two of the three wavelength regions are not used because
they lacked sufficient non-flagged pixels. At $z=1$ each point on the
plot includes pixels from a $z$ range of 0.0548, hence all three
points from each QSO appear in the same part of the plot. The relative
lack of points at z=0.3 and 0.95 is caused by the $z$ distribution of
the QSOs in the HST archive and the SNR in the \lyaf\ of those
spectra.  The negative DA values show a combination of the true amount
of absorption, the continuum level errors and photon noise.  The large
dispersion in the points at a given $z$ is largely real variability in
the IGM. The occasional very high points typically have two
conspicuous deep lines that were not masked.  The mean absorption
apparently increases $z$.

\begin{figure}
   \includegraphics[width=84mm]{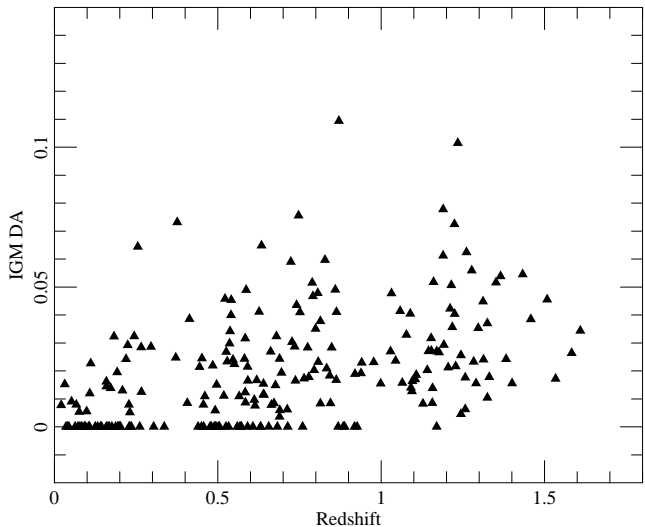}
   \caption{\label{fosonlyw} As Fig. \ref{fosonlyda} for the same
     QSOs and spectra but calculated by summing the equivalent widths of line
     listed in B02.  This method gives less absorption that we
     measure from the flux in each pixel.
     }
\end{figure}

We also show a power law ${\rm DA}(z) = A(1+z)^{\alpha}$ which is a
fit to these points, not to the DA per pixel. We found $A = 0.016$ and
$\alpha = 1.01$.  We will not give errors on $A$ and $\alpha $ because
the two parameters are strongly anti-correlated.  For fixed $A=0.016$,
$\sigma _{ \alpha } = 0.11$, and for fixed $\alpha = 1.01$, $\sigma _{
  A } = 0.001$. However it is not unusual to see the value of $\alpha
$ change by many tenths for common changes in the sample of spectra.

In Table \ref{tablefosda} we list values obtained from the points
shown in Figure \ref{fosonlyda}, binned into widths of $\delta z =
0.2$, where the first bin covers $0.1 < z < 0.3$. The column labeled
``IGM'' shows the total DA from the low density IGM alone, that is with
the metals and strong H~I lines masked.  The column $\sigma_{\rm IGM}$
is the standard deviation of the points in the bin, divided by the
square root of the number of points (the standard error).  The column
``All H'' includes pixels flagged as H~I lines, while the column
labeled ``Total'' includes all pixels not flagged as Galactic
absorption.  The column labeled ``Metal'' uses only those pixels
flagged as metal lines, column labeled ``IDH'' refers to the strong
lines that were Identified as H by B02.  The values given in the
column ``All H'' are defined as the sum of IGM + IDH.  The IDH values
are approximately constant with $z$, with a mean of 0.0032.  The metal
absorption has a mean value of 0.0045 averaged over all redshifts in
the FOS data, and it systematically increases with increasing
redshift.

When we ignore redshift and take the mean for the
whole sample, we find the total absorption of DA = 0.0371 is comprised of 
0.0293 (79\%) low density IGM,
0.0043 (12\%) metals, and
0.0034 (9\%) strong H~I lines.
The signal is dominated by the low density IGM, just as we found at
$z = 2$ where the values were 78, 15 and 7\% \citep{tytler04a}.

The $\sigma_{\rm IGM}$ is a representation of the error on the DA that
we list for the IGM alone.  The external error is probably larger,
because the points are not fully independent; there are large scale
correlations in the amount of absorption across spectra, and
systematic errors such as continuum level placement also extend across
spectra.

The external error is also larger because the continuum level errors
are partly but not completely contained in the errors that we quote in
Table \ref{tablefosda}.  The relative difference in IGM DA for our two
continuum fits (that is, the difference in our DA values at each
redshift divided by the mean DA at that redshift) has a root mean
squared value of 0.21.  This is about the same size as the
$\sigma_{\rm IGM}$ that we list in Table \ref{tablefosda}, which we
take to mean that our errors due to continuum fitting and the (Poisson
like) uncertainty due to our sampling of large scale structure are
similar in magnitude.  Immediately prior to finalizing our continuum
fits, we had two relatively independent fits, that gave mean DA values
averaged over all z, and these two DA values differed by a factor of
1.049.  Earlier in the fitting process, the two continua differed by
much larger amounts, for reasons that were apparent and which were
taken into account as we iterated on the continuum fitting.  The
values given here are far from a precise measurements of the external
errors, but they may indicate the approximate size of the error from
continuum fitting alone.  We speculate that the external errors for
the IGM DA are of order 1.4 to 2 times the values given in Table
\ref{tablefosda} for redshift intervals of 0.2.

The effect of a given change in the continuum level on the DA value
depends on the distribution of the absorption line depths. The lower
the spectral resolution, the shallower the lines, and the more
sensitive the DA to a given continuum level change.  If the evolution
of the DA with $z$ is primarily a change in the number of lines and
not in the line depths, then a given change in continuum level will
produce the same relative change in DA at all $z$.  We know, for
example, that strong (typically deep) lines are increasingly common at
higher $z$ \citep{janknecht06a}, which will make DA less sensitive
to a given fractional error in the continuum.

\begin{table*}
\caption{\label{tablefosda} DA as a function of redshift from rest frame wavelengths
1070 -- 1170~\AA\ in FOS spectra of the 74 QSOs}
\begin{tabular}{l cc cc cc cc cc}
\hline
$z$  & IGM   &  $\sigma_{\rm IGM}$  & All H  & $\sigma_{\rm All H}$  
     &  Total & $\sigma_{\rm Total}$ &  Metal & $\sigma_{\rm Metal}$ 
     & IDH & $\sigma_{\rm IDH}$ \\
\hline
0.20 & 0.020 & 0.003 &   0.023 & 0.003 &  0.024 & 0.003 & 0.000 & 0.000 & 0.003 & 0.003\cr
0.40 & 0.020 & 0.003 &   0.022 & 0.003 &  0.024 & 0.003 & 0.002 & 0.002 & 0.003 & 0.002\cr
0.60 & 0.026 & 0.003 &   0.028 & 0.003 &  0.031 & 0.003 & 0.003 & 0.001 & 0.003 & 0.001\cr
0.80 & 0.035 & 0.004 &   0.042 & 0.005 &  0.045 & 0.005 & 0.003 & 0.001 & 0.007 & 0.002\cr
1.00 & 0.037 & 0.004 &   0.040 & 0.004 &  0.049 & 0.006 & 0.010 & 0.004 & 0.003 & 0.002\cr
1.20 & 0.045 & 0.005 &   0.047 & 0.005 &  0.057 & 0.005 & 0.010 & 0.003 & 0.002 & 0.002\cr
1.40 & 0.039 & 0.007 &   0.043 & 0.010 &  0.051 & 0.013 & 0.008 & 0.004 & 0.004 & 0.004\cr
1.60 & 0.027 & 0.007 &   0.027 & 0.007 &  0.044 & 0.012 & 0.017 & 0.007 & 0.000 & -0.000\cr

\end{tabular}
\end{table*}

\begin{figure}
  \includegraphics[width=74mm]{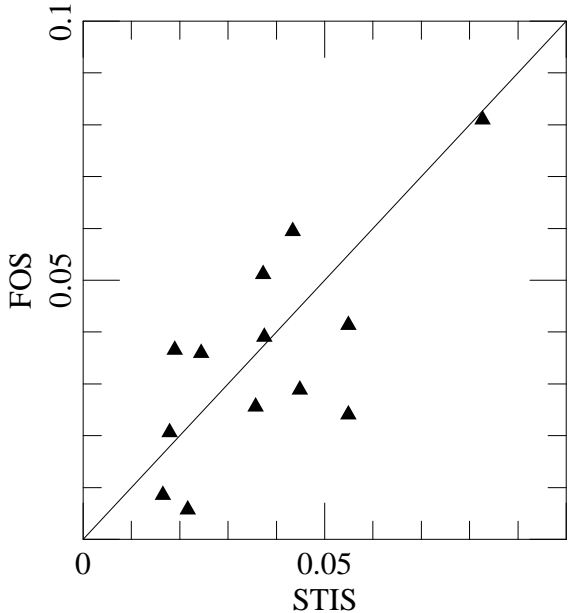}
  \caption{\label{stisda}The DA from FOS spectra versus the DA from
    STIS spectra of the same objects. Five QSOs each contribute up to
    three points, each point using the flux from 33.3~\AA\ in the rest
    frame. Three segments are not shown because more than 30\% of their
    pixels were masked. The solid diagonal line shows expectation for ideal data and
    measurements.  }
\end{figure}

\begin{figure}
   \includegraphics[width=84mm]{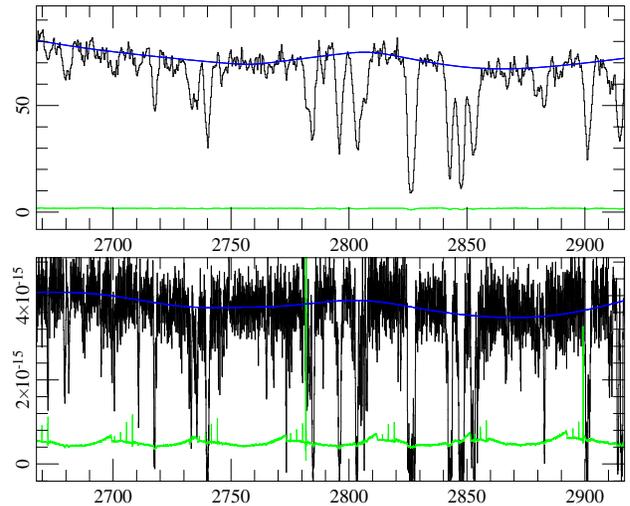}
   \caption{\label{pgstis} Rest frame wavelengths 1070 -- 1170~\AA\
in the FOS (top) and STIS (bottom) spectra of QSO 
PG 0117+213 at \zem = 1.493. The upper smooth curve is the continuum while the
lower line just above zero is the $1\sigma $ error. 
Known emission lines that vary in strength from QSO to QSO cause the
continuum level to rise at either end and near the middle.
}
\end{figure}

The five QSOs indicated in Table \ref{fosspectab} have STIS as well as
FOS spectra.  In Figure \ref{stisda} we compare the DA measured from
both spectra, in our three wavelength bins.  This shows that our
continuum placements and masks yield similar DA values, irrespective
of the instrument.  The masks for the STIS spectra are from
\citet{milutinovic06a} while those for the FOS spectra are from
B02. In Figure \ref{pgstis} we show the two spectra for PG 0117 which
has intermediate SNR STIS and FOS spectra.  We would expect the
continua on the STIS spectra to be more accurate, since more photons
are recorded per spectrum, while \citet{milutinovic06a} have made a
concerted effort to make the metal line identifications complete.

We also calculate a version of DA for the low density IGM by summing
the W values that B02 give for the unidentified lines in the spectra
of the 74 QSOs. We measure this DA in the same redshift bins that we
used in Fig. \ref{fosonlyda}. In each bin we obtain DA by subtracting
the observed frame W values from the observed frame path length in \AA
ngstroms, and dividing the result by the path length.  The results in
Fig. \ref{fosonlyw} show general similarity with the DA defined from
the flux in Fig. \ref{fosonlyda} except that the flux DA shows more
absorption, as we might expect. When we use the flux in pixels to
measure DA, we can detect absorption from features that do not make
the threshold to count as reliable lines. For these FOS spectra, the
line threshold varies from spectrum to spectrum, following the SNR.
When we average over all $z$, the two DA values for the low density
IGM are 0.020 from the lines, and 0.029 from the flux, larger by a
factor of 1.46.

Following \citet{weymann98},
\citet{dobrzycki02} fit the number of lines in the 336 B02 spectra with a
power law in redshift and an exponential in rest equivalent width
\begin{equation}
{\partial^2 N \over \partial z \partial W_{\lambda0}} =
    A_0  W_*^{-1} (1+z)^\gamma e^{-W_{\lambda0}/W_*},
\end{equation}
where $W_{\lambda0}$ is the rest equivalent width of a line, and 
$A_0$, $W_*$, and $\gamma$ are constants that are fit to match the
observed distribution of line $W_{\lambda0}$.  The mean of the exponential
distribution $e^{-(W_{\lambda0}/W_*)}$ from zero to infinity 
is $W_*$, so the mean of $W_*^{-1} e^{-(W_{\lambda0}/W_*)}$ is unity, 
and the number density of lines as a function of redshift is 
\begin{equation}
{\partial N \over \partial z} = A_0 (1+z)^\gamma .
\end{equation}

Since the average observed equivalent with of a line at redshift $z$
is $W=W_* (1+z)$, we find that the mean observed frame 
equivalent width of absorption as a function of redshift is, as usual, given by
\begin{equation}
{\partial W \over \partial z} = W_* (1+z) {\partial N \over \partial z}
   = A_0 W_* (1+z)^{\gamma + 1}.
\end{equation}
To convert to DA$(z)$, all we need to do is divide by the available 
observed frame path length per unit redshift, 
$\partial \lambda / \partial z = \lambda_0$
where $\lambda_0$ is the rest frame wavelength of the absorption line,
or 1215.67 \AA\ for H~I \lya.  This gives
\begin{equation}
{\rm DA}(z) = {A_0 W_* \over \lambda_0} (1+z)^{\gamma + 1}.
\end{equation}
This shows that if the rest frame equivalent width distribution is 
independent of redshift
and the number of lines per unit z changes as $(1+z)^{\gamma }$, then
DA$(z) \propto (1+z)^{\gamma +1}$, where the extra one in the power comes
from the increase in the observed equivalent width with $1+z$.

The exponent $\alpha = 1 + \gamma = 1.01$ that we measured for the
flux in the pixels in the FOS spectra is consistent with $\gamma = 0$
which means no change in the number lines per unit redshift, if the
lines have fixed rest frame equivalent widths.  From Table 2 in
\citet{dobrzycki02} we see that the evolution of \lyaf\ lines in the
FOS data is characterized by $1+\gamma \sim 1.5$, larger than the
value that we measured. Before we comment on this difference, we
should look at the amplitudes.

For the ``forest'' sample with an equivalent width threshold of 0.24
\AA, and no maximum $W$ cutoff (line 6 of Table 2 in
\citet{dobrzycki02}), the amplitude is $A_0 W_* / \lambda_0 = 0.0048$,
about one third of what we find.  But the line density $A_0$ in
\citet{dobrzycki02} is defined to give the density of lines with $W >
0.24$ \AA . To compare with DA we can alternatively extrapolate to
$W=0$, which increases their amplitude by a factor of $
\int^{\infty}_0 e^{-W_{\lambda_0}/W_*} / \int^{\infty}_{0.24}
e^{-W_{\lambda_0}/W_*} = 2.9$ when $W_* = 0.22$ \AA.  This
extrapolation gives an amplitude of 0.0139, similar to the 0.014 that
we found from the flux in pixels.  This agreement is slightly
misleading because they include and we exclude strong H~I lines.

It is possible to have both $1 + \gamma = 1.0$ for the DA from flux in
pixels, and $1 + \gamma = 1.5$ for the lines with $W > 0.24$ \AA ,
since the latter includes only 1/3 of the absorption in the former.
This would require that the stronger lines, the only ones seen in line
counting, decline rapidly in number with decreasing redshift. The
weaker lines that dominate the total absorption that we measure using
the flux in the pixels change little.  The more rapid evolution of the
stronger lines has been reported before.  \citet{janknecht06a} found
that lines with \lnhi\ $> 13.64$~\cmm\ had $1+ \gamma = 2.50 \pm 0.45$
while lines with $12.9 < $ \lnhi\ $< 14.0$~\cmm\ had $1+ \gamma = 0.74
\pm 0.31$, both for $0.7 < z < 1.9$.  \citet{janknecht06a} also note
that the evolution is near zero at lower redshifts for even stronger
lines. They find $1 + \gamma = 1.13 \pm 0.06$ for the number of lines
with $W > 0.24$~\AA\ at $0 < z < 1.5$ from \citet{weymann98}.

\section{FOS Flux PDF and Autocorrelation}

\begin{figure*}
  \includegraphics[width=100mm, angle=270]{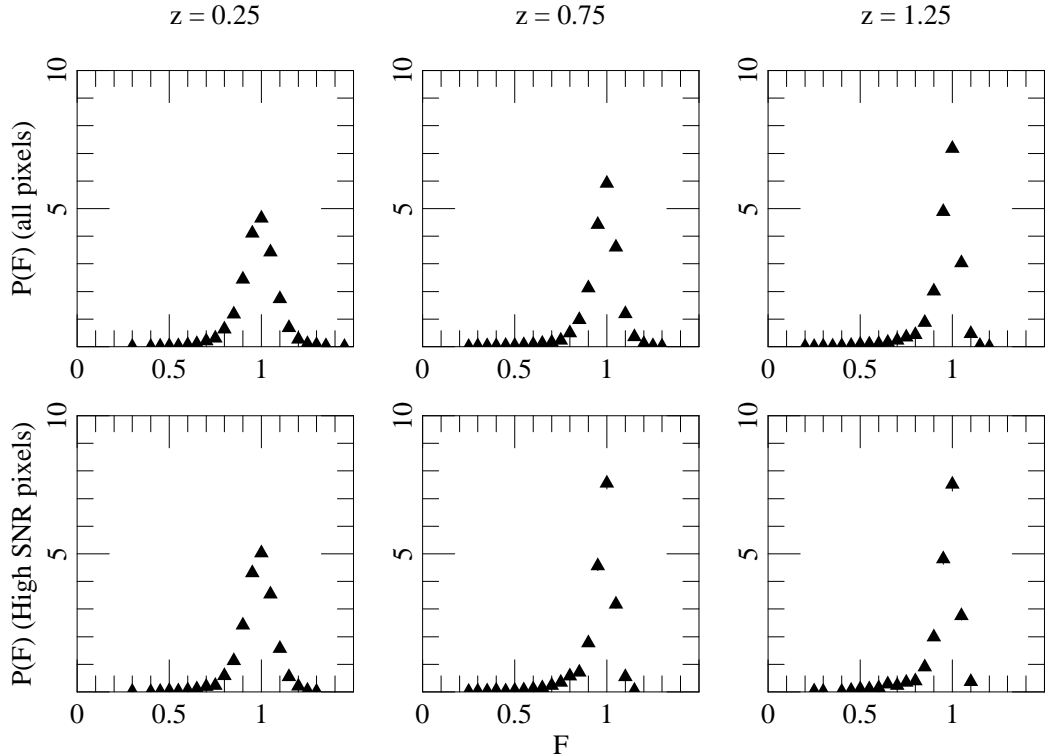}
  \caption{\label{fluxpdf} PDF of the flux in FOS spectra of the 74 QSO, 
    from rest frame wavelengths 1070 -- 1170~\AA . The PDF is evaluated in three
    independent redshift bins. There are 4 pixels per FOS diode.
    We masked pixels containing metal or strong \lya\ lines, and hence these plots
    should contain absorption from only the low density IGM.
    The lower panels use the pixels with
    SNR $> 40$ and the higher panels use pixels with all SNR values.}
\end{figure*}

\begin{table*}
\caption{\label{tablefluxpdf} PDF of the flux at 1070 -- 1170~\AA\
  in the FOS spectra of the 74 QSO,
  for the low density IGM pixels only. We have masked all the strong \lya\
  and metal lines.  }
\begin{tabular}{c cc cc cc cc cc cc cc}
\hline
      &   \multicolumn{4}{c}{$0 < z <0.5$}
      &   \multicolumn{4}{c}{$0.5 < z <1$}
      &   \multicolumn{4}{c}{$1 < z <1.5$}
\cr
      &   \multicolumn{2}{c}{All}  & \multicolumn{2}{c}{High SNR}
      &   \multicolumn{2}{c}{All}  & \multicolumn{2}{c}{High SNR}
      &   \multicolumn{2}{c}{All}  & \multicolumn{2}{c}{High SNR}
\cr
 F       & $P(F)$ & $\sigma_{P(F)}$
         & $P(F)$ & $\sigma_{P(F)}$
         & $P(F)$ & $\sigma_{P(F)}$
         & $P(F)$ & $\sigma_{P(F)}$
         & $P(F)$ & $\sigma_{P(F)}$
         & $P(F)$ & $\sigma_{P(F)}$
\cr
\hline
0.050 & 0.00 & 0.00 & 0.00 & 0.00 & 0.00 & 0.00 & 0.00 & 0.00 & 0.00 & 0.00 & 0.00 & 0.00 \cr
0.100 & 0.00 & 0.00 & 0.00 & 0.00 & 0.00 & 0.00 & 0.00 & 0.00 & 0.00 & 0.00 & 0.00 & 0.00 \cr
0.150 & 0.00 & 0.00 & 0.00 & 0.00 & 0.00 & 0.00 & 0.00 & 0.00 & 0.00 & 0.00 & 0.00 & 0.00 \cr
0.200 & 0.00 & 0.00 & 0.00 & 0.00 & 0.00 & 0.00 & 0.00 & 0.00 & 0.00 & 0.00 & 0.00 & 0.00 \cr
0.250 & 0.00 & 0.00 & 0.00 & 0.00 & 0.00 & 0.00 & 0.01 & 0.01 & 0.01 & 0.00 & 0.02 & 0.01 \cr
0.300 & 0.01 & 0.00 & 0.01 & 0.01 & 0.02 & 0.00 & 0.03 & 0.01 & 0.01 & 0.00 & 0.02 & 0.01 \cr
0.350 & 0.00 & 0.00 & 0.00 & 0.00 & 0.02 & 0.00 & 0.02 & 0.01 & 0.01 & 0.00 & 0.00 & 0.00 \cr
0.400 & 0.00 & 0.00 & 0.01 & 0.00 & 0.03 & 0.01 & 0.06 & 0.02 & 0.03 & 0.01 & 0.02 & 0.01 \cr
0.450 & 0.01 & 0.00 & 0.01 & 0.01 & 0.04 & 0.01 & 0.03 & 0.01 & 0.04 & 0.01 & 0.05 & 0.02 \cr
0.500 & 0.02 & 0.01 & 0.03 & 0.01 & 0.04 & 0.01 & 0.05 & 0.02 & 0.08 & 0.01 & 0.11 & 0.03 \cr
0.550 & 0.03 & 0.01 & 0.02 & 0.01 & 0.06 & 0.01 & 0.07 & 0.02 & 0.09 & 0.01 & 0.10 & 0.03 \cr
0.600 & 0.06 & 0.01 & 0.06 & 0.01 & 0.08 & 0.01 & 0.11 & 0.02 & 0.10 & 0.01 & 0.15 & 0.03 \cr
0.650 & 0.10 & 0.01 & 0.10 & 0.02 & 0.11 & 0.01 & 0.14 & 0.03 & 0.16 & 0.01 & 0.28 & 0.05 \cr
0.700 & 0.20 & 0.02 & 0.19 & 0.02 & 0.15 & 0.01 & 0.23 & 0.04 & 0.23 & 0.02 & 0.22 & 0.04 \cr
0.750 & 0.30 & 0.02 & 0.22 & 0.03 & 0.23 & 0.02 & 0.34 & 0.04 & 0.34 & 0.02 & 0.34 & 0.05 \cr
0.800 & 0.63 & 0.03 & 0.58 & 0.04 & 0.49 & 0.02 & 0.57 & 0.06 & 0.43 & 0.02 & 0.39 & 0.05 \cr
0.850 & 1.17 & 0.05 & 1.12 & 0.06 & 0.97 & 0.03 & 0.71 & 0.06 & 0.87 & 0.03 & 0.89 & 0.08 \cr
0.900 & 2.44 & 0.07 & 2.41 & 0.08 & 2.13 & 0.05 & 1.77 & 0.10 & 2.01 & 0.05 & 1.98 & 0.12 \cr
0.950 & 4.10 & 0.09 & 4.31 & 0.11 & 4.42 & 0.07 & 4.56 & 0.16 & 4.88 & 0.08 & 4.81 & 0.19 \cr
1.000 & 4.65 & 0.09 & 5.02 & 0.12 & 5.91 & 0.08 & 7.55 & 0.21 & 7.17 & 0.10 & 7.51 & 0.24 \cr
1.050 & 3.42 & 0.08 & 3.53 & 0.10 & 3.60 & 0.06 & 3.17 & 0.13 & 3.03 & 0.06 & 2.75 & 0.14 \cr
1.100 & 1.73 & 0.06 & 1.57 & 0.07 & 1.19 & 0.03 & 0.54 & 0.06 & 0.47 & 0.03 & 0.37 & 0.05 \cr
1.150 & 0.68 & 0.04 & 0.55 & 0.04 & 0.35 & 0.02 & 0.05 & 0.02 & 0.04 & 0.01 & 0.00 & 0.00 \cr
1.200 & 0.27 & 0.02 & 0.20 & 0.02 & 0.10 & 0.01 & 0.00 & 0.00 & 0.00 & 0.00 & 0.00 & 0.00 \cr
1.250 & 0.11 & 0.01 & 0.06 & 0.01 & 0.03 & 0.01 & 0.00 & 0.00 & 0.00 & 0.00 & 0.00 & 0.00 \cr
1.300 & 0.06 & 0.01 & 0.01 & 0.01 & 0.01 & 0.00 & 0.00 & 0.00 & 0.00 & 0.00 & 0.00 & 0.00 \cr
1.350 & 0.02 & 0.01 & 0.00 & 0.00 & 0.00 & 0.00 & 0.00 & 0.00 & 0.00 & 0.00 & 0.00 & 0.00 \cr
1.400 & 0.00 & 0.00 & 0.00 & 0.00 & 0.00 & 0.00 & 0.00 & 0.00 & 0.00 & 0.00 & 0.00 & 0.00 \cr
1.450 & 0.00 & 0.00 & 0.00 & 0.00 & 0.00 & 0.00 & 0.00 & 0.00 & 0.00 & 0.00 & 0.00 & 0.00 \cr
1.500 & 0.00 & 0.00 & 0.00 & 0.00 & 0.00 & 0.00 & 0.00 & 0.00 & 0.00 & 0.00 & 0.00 & 0.00 \cr

\end{tabular}
\end{table*}
  
\begin{figure*}
  \includegraphics[width=60mm, angle=270]{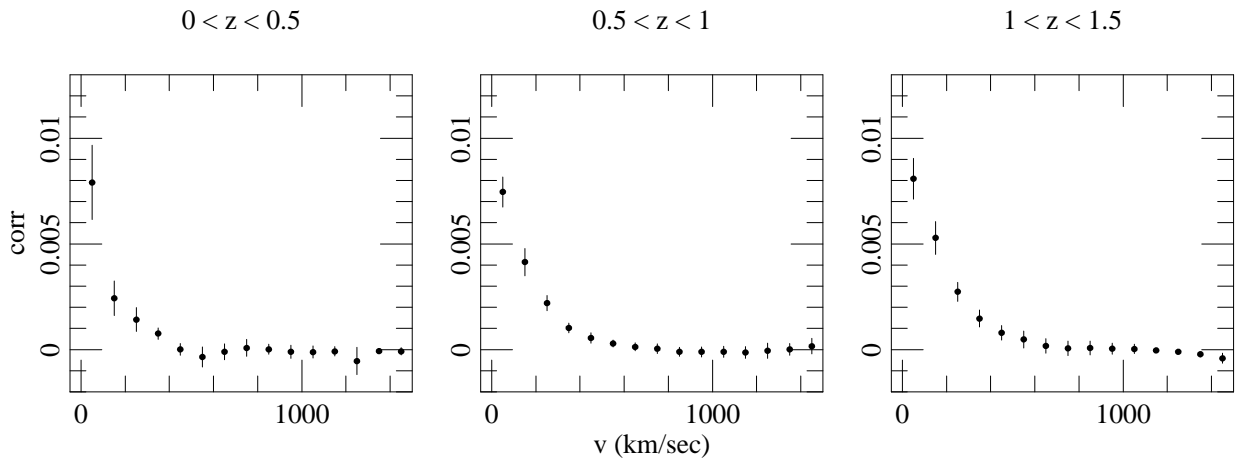}
  \caption{\label{figauto} As Fig. \ref{fluxpdf} but showing the autocorrelation 
   of the flux defined in Eqn. \ref{eqncorr} as a function of velocity lag.
}
\end{figure*}

\begin{table*}
\caption{\label{tableauto} 
  As Table \ref{tablefluxpdf}, but for the flux autocorrelation defined in
  Eqn. \ref{eqncorr} as a function of the velocity in the center of
  the 100~\kms\ wide bins. 
 }
\begin{tabular}{r ccc ccc}
\hline
$v$  &   \multicolumn{2}{c}{$0 < z <0.5$}
     &   \multicolumn{2}{c}{$0.5 < z <1$}
     &   \multicolumn{2}{c}{$1 < z <1.5$}
\cr
(km/sec) 
     & $\xi$ & $\sigma_\xi$ 
     & $\xi$ & $\sigma_\xi$
     & $\xi$ & $\sigma_\xi$
\cr
\hline
 50.0  &   0.007906&  0.001374 &  0.007456  &  0.000771 &  0.008080 &  0.001014 \cr
 150.0  &   0.002431&  0.000626 &  0.004138  &  0.000653 &  0.005285 &  0.000791 \cr
 250.0  &   0.001419&  0.000504 &  0.002193  &  0.000354 &  0.002733 &  0.000510 \cr
 350.0  &   0.000760&  0.000316 &  0.001022  &  0.000256 &  0.001471 &  0.000439 \cr
 450.0  &   0.000009&  0.000539 &  0.000544  &  0.000209 &  0.000788 &  0.000304 \cr
 550.0  &  -0.000346&  0.000426 &  0.000287  &  0.000149 &  0.000480 &  0.000425 \cr
 650.0  &  -0.000104&  0.000384 &  0.000126  &  0.000207 &  0.000173 &  0.000327 \cr
 750.0  &   0.000082&  0.000335 &  0.000044  &  0.000196 &  0.000062 &  0.000314 \cr
 850.0  &   0.000012&  0.000316 & -0.000104  &  0.000205 &  0.000081 &  0.000279 \cr
 950.0  &  -0.000103&  0.000300 & -0.000106  &  0.000205 &  0.000040 &  0.000268 \cr

\end{tabular}
\end{table*}

We also calculate the probability distribution of the flux -- the flux
PDF -- and the autocorrelation of the flux, two of the more common
higher order continuous statistics. We use exactly the same spectra
and masking to leave only the low density IGM that we used for the DA
calculation. The center of mass of the PDF should be the DA
value. This is unfortunately the only information that we can readily
obtain from the PDF without knowing the spectral resolution
and SNR for each pixel.

In Figure \ref{fluxpdf} and Table \ref{tablefluxpdf} we show the PDF
in six panels, divided at redshifts 0.5 and 1.0 and showing the high
SNR spectra alone.  P(F) is normalized so that its integral over all F
is unity.  Some pixels have flux F above 1.0 because of a combination
of photon noise and continuum level errors.  We believe that our
continuum level errors are smaller in units of F when the SNR is
higher. The lower three panels use only pixels with SNR $> 40$. They
show slightly fewer pixels with $F > 1.0$.

We see that the most likely F is in the range 0.975 -- 1.025.  This
is expected if most pixels in the FOS spectra have $DA < 0.025$.  If
most pixels have $DA < 0.025$ then the mean error in our continuum
level is $< 0.025$. Alternatively if most pixels have $DA > 0.025$ then
the continuum error is large and unknown. At these low redshifts
visible absorption lines occupy a relatively small part of each
spectrum. If most of the total DA is in lines and not in a smooth GP
absorption then we expect most pixels to be near the continuum.

The second most common flux value is that immediately below F=1.0, as
we expect for absorption lines. The flux PDF in the lowest redshift
bin is similar to a Gaussian, centered just below F=1.0, while at
higher $z$ we see a tail of lower F values from deeper lines. We do
not see many pixels with F near zero. This is mostly because of the
low spectral resolution, and also because we mask the strongest lines.
If we use high resolution spectra that resolve all lines, most with
column densities $ 14 < $ \lnhi $ < 17$~\cmm\ would reach near zero
flux, and some would not be masked using the criteria adopted here.
Using only the higher SNR pixels gives somewhat narrower
distributions, as we would expect, but this is a small change because
all the spectra have relatively high SNR.

We calculated the autocorrelation of the flux in the 74 spectra in the
interval 1070 -- 1170~\AA\ with the usual masks to leave only the low
density IGM. We use a common definition for the autocorrelation at
velocity lag $v$
\begin{equation}
\label{eqncorr}
\xi(v) = 
\left<
(F_i -\overline{F} )
(F_{i + v}-\overline{F} )
\right>
\end{equation}
where $F_{i+v}$ is the flux in a pixel separated from pixel $i$ by a
velocity $v$ and $\overline{F}$ is the mean flux in each spectrum, which
differs from spectrum to spectrum.  To
estimate the uncertainty in $\xi(v)$, we divided each redshift
interval into 16 separate samples, and took the standard deviation of
the sub-samples.

In Figure \ref{figauto} and Table \ref{tableauto} we show the results.
The autocorrelation is non-zero to $>500$~\kms\ ($\sim 1\sigma $) for
all three redshift intervals, and exceeds $2\sigma $ out to $\sim
400$~\kms\ in lowest redshift bin and to $\sim 500$~\kms\ in the other
two bins. Overall the autocorrelation is larger at higher
redshifts. The velocity width of the signal is comparable to, but
wider than expected from the spectral resolution. We see only marginal
correlation beyond the spectral resolution and this is consistent with
results at higher redshifts where high resolution spectra show little
correlation beyond $150$~\kms . The points in each plot are highly
correlated by construction, and hence it is unclear how much
significance we should assign to the apparent broader distribution at
higher redshifts.  \citet{janknecht06a} saw $2\sigma $ correlation out
to $< 200$~\kms\ for lower column density lines and to $< 100$~\kms\
for higher column density lines. The FOS spectra that we use here have
more correlation at intermediate lags $\sim 100$~\kms , the $\sigma $
of the Gaussian representing the line spread function.  On larger
scales the spectral resolution has limited effect, and the FOS spectra
remain more sensitive because they contain 74 QSO spectra compared to
the 9 STIS spectra used by \citet{janknecht06a}.

\section{IGM H I opacity from $0 < z < 3.2$}

\begin{figure}
   \includegraphics[width=84mm]{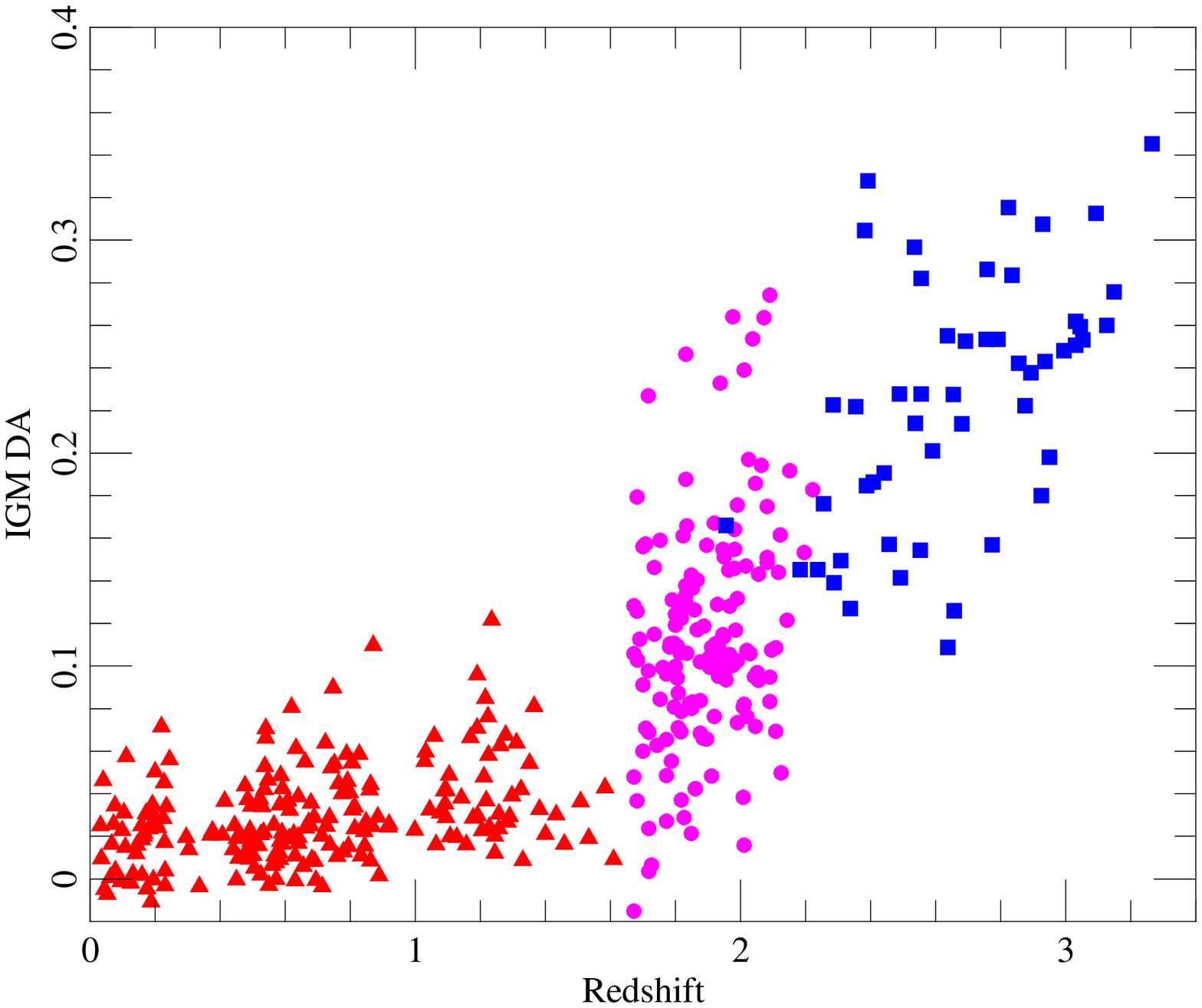}
   \caption{\label{allda} The DA as a function of redshift measured in
     QSO spectra obtained with three spectrographs: FOS (triangles) at
     the low redshifts, Kast (circles) at intermediate redshifts and
     HIRES (dark squares) at the highest redshifts. Each point from
     FOS represents 33.3~\AA\ in the rest frame, as in earlier
     figures, while the other points are for 121.56~\AA\ in the
     observed frame from \citet{kirkman05a}. We mask the identified
     strong \lya\ and metals lines in the FOS and HIRES spectra, but
     we subtract the mean expected amounts of such and metal line
     absorption from all the Kast points.  }
\end{figure}

In Figure \ref{allda} we show the DA over a range of redshifts from
the FOS data discussed here, from our Kast spectra \citep{tytler04a}
and from our HIRES spectra \citep{kirkman05a}. An extrapolation of the
best fitting power law from $1.6 < z < 3.2$ gives more absorption than
we see at $z \sim 1.5$, but then less at the lowest redshifts. The
power law fit to $z < 1.6$ alone crosses that from higher $z$ at $z =
0.7$.

In Table \ref{tabledabin} we bin the DA values shown Fig. \ref{allda}
in redshift intervals of 0.2. We list the mean DA in each bin and its
error.  The best fit single power law to the DA values in Table
\ref{tabledabin} from FOS, Kast and HIRES gives $A = 0.0066$ and
$\alpha = 2.661$, but with an unacceptably large $\chi^2 = 41.1$ for
14 degrees of freedom. The fit improves significantly when we use a
broken power law:
\begin{equation}
\label{eqnpower}
{\rm DA}(z) = \left\{ A (1+z)^{\alpha_l} \quad : \quad z < z_c  \atop
                      B (1+z)^{\alpha_h} \quad : \quad z \ge z_c
              \right.
\end{equation}
where $B = A (1+z_c)^{\alpha_l} / (1+z_c)^{\alpha_h}$, and $A$, $z_c$, $\alpha_l$,
and $\alpha_h$ are free parameters.  The best fit we can find ($\chi^2 = 22.1$
for 12 degrees of freedom, Prob($\chi ^2 > 22.1) = 4$\%),
is not an excellent representation of the data. 
This fit has $A = 0.013$, $z_c = 1.1$, $\alpha_l = 1.54$, and $\alpha_h = 2.98$.
\citet{janknecht06a} found that the distribution of lines at $0.5 < z < 1.9$
were fit with $\alpha = 1.74 \pm 0.31$ from STIS spectra of nine QSOs,
consistent with our $\alpha_l$ value.

\begin{figure}
   \includegraphics[width=84mm]{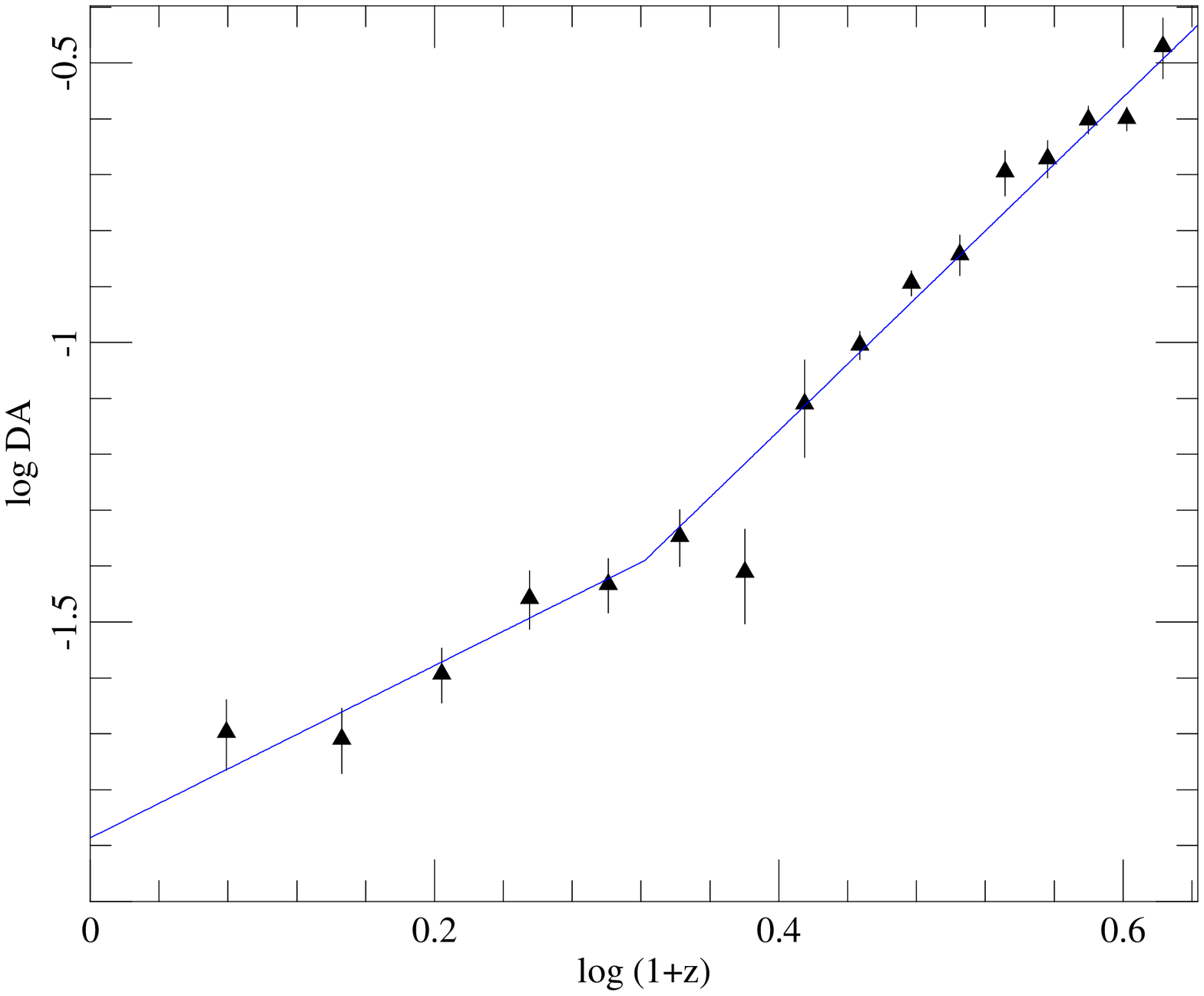}
   \caption{\label{logdafit} log DA as a function of log(1+z) from
     Table \ref{tabledabin}. The solid line is the best fit broken power
     law of Eqn. \ref{eqnpower}.  }
\end{figure}

In Figure \ref{logdafit} we show the points from Table
\ref{tabledabin} and the broken power law.  Although the data seem to
depart from a simple smooth distribution at several redshift, this is
probably an indication that we have underestimated the errors.  The
data are noticeably lower than the fit at $z=1.4$, for no known
reason.  This is the highest redshift bin that uses FOS spectra alone,
and it samples redshifts where we have few spectra.

\begin{table}
\caption{\label{tabledabin}DA as function of redshift from FOS, Kast and HIRES}
\begin{center}
\begin{tabular}{llll}
\hline
$z$ & DA & $\sigma_{\rm DA}$  \cr
\hline
0.20 & 0.020 & 0.003  \cr
0.40 & 0.020 & 0.003  \cr
0.60 & 0.026 & 0.003  \cr
0.80 & 0.035 & 0.004  \cr
1.00 & 0.037 & 0.004  \cr
1.20 & 0.045 & 0.005  \cr
1.40 & 0.039 & 0.007  \cr
1.60 & 0.078 & 0.015  \cr
1.80 & 0.099 & 0.006  \cr
2.00 & 0.128 & 0.007  \cr
2.20 & 0.144 & 0.012  \cr
2.40 & 0.202 & 0.019  \cr
2.60 & 0.213 & 0.016  \cr
2.80 & 0.250 & 0.014  \cr
3.00 & 0.251 & 0.012  \cr
3.20 & 0.338 & 0.042  \cr

\end{tabular}
\end{center}
\end{table}

\section{Discussion and Conclusion}
\label{discuss}

We have presented common continuous statistics on the H~I absorption
in the low density IGM at $0 < z < 1.6$. We work mostly with the flux
in each pixel, but also with line lists and fits to the distributions
of lines. 

DA(z) is a function that we have
previously used at higher redshifts to deduce the physical properties
of the IGM \citep{tytler04b}. We generally require that simulations of
the IGM should match the observed DA when they are using a concordant
set of cosmological and astrophysical parameters. When these other
parameters are known from other work, we can use DA to give the most
accurate measurements of the intensity of the UV background radiation
that ionizes the IGM.

The DA is an explicit and easy to understand measurement of the total
absorption by \lya\ in the IGM. It is relatively easy to visualize the
effects of the continuum level error and contaminants.  DA, as a mean,
is the only moment of the flux PDF that does not explicitly depend on spectral
resolution or SNR, although both do effect the continuum level placement
and line identification. This makes DA easy to use, especially with spectra
of low, mixed or poorly known resolution.

The redshift evolution of DA is somewhat slower that given by published
analysis of the distribution of individual lines.  From the DA
measured from the flux in pixels, we find $\gamma +1= 1.01$ while
\citet{dobrzycki02} find $1+ \gamma \sim 1.5 \pm 0.2$ from fits to the
number of lines in all 336 B02 spectra.
The difference may be explained by the more rapid evolution of the stronger 
lines used by \citet{dobrzycki02}, lines which constitute only 1/3 of the 
total absorption counted by the flux in the pixels.

The DA for the low density IGM that we measure from the flux in each
pixel contains 1.46 times more absorption than is contained in lines
seen in the same spectra.  This is less than the factor of 3
difference from the larger sample of 336 FOS QSOs, because the large
sample is does not have a uniform minimum rest frame $W$ value, and it
includes spectra with lower SNR -- spectra that are less sensitive to
weak lines.  However, if we fit the distribution of line W values with
an exponential function and we extrapolate this fit to zero W, we
recover most of the missing absorption. The exponential distribution
gives approximately the correct amount of absorption that is not in
lines seen individually in these FOS spectra.  These results are
broadly consistent with those from higher redshifts.  At $z=2.7$
\citet{kirkman97a} found that only half of the absorption came from
lines with $\lnhi > 14~\cmm$ which are the rare saturated lines easily
seen in FOS spectra.

The main limitation in line counting is the inability to individually
identify lines with low W. Measurement of DA from flux can include
these lines, but they are mostly shallower, and hence more sensitive
to the error in the placement of the continuum. The continuum level
error is the main error in the DA value, because much of the
absorption is from shallow features. This sensitivity to the continuum
level is intrinsic to the absorption, and does not depend on how we
measure the absorption, from the flux per pixel or from line lists.

We found in \citet{tytler04b} that we required high SNR spectra to
obtain accurate DA values. Here we also need high SNR to give
continuum level errors that are not much larger than the DA signal. We 
limit our measurements to the relatively unusual FOS spectra with SNR
$> 20$ per pixel.

In addition to the SNR, the continuum error also depends on the the
emission lines, the range of wavelengths in a spectrum and the
methods used. We fit the continuum by hand because we do not yet have
automatic methods that do better. \citet{suzuki05a} used principal
component spectra trained on HST spectra with absorption removed. They
then attempted to predict the continuum in the \lyaf\ from the
components that fit the red side of a spectrum, where the absorption
is rare.  This was only partly successful, we suspect, because of 
changes in slope (intrinsic, and from Galactic and atmospheric extinction)
across each spectrum.

The emission lines in the \lyaf\ are of different strengths in
different QSO spectra. The continuum is harder to fit when the lines are
strong (e.g. 0102-2713), especially in low SNR spectra with abundant
absorption. We made thin tall plots aligned in rest wavelength to help
us find these lines, and make their shapes consistent with the other
lines in the spectrum.  If we underestimate the strengths of the
emission lines, which is readily done, then we will tend to place the
continuum too low near those lines, giving systematically too little
absorption at their wavelengths.  It is very important to see a large
portion of each spectrum, well beyond the region of interest for DA,
both to establish the general strengths and shapes of the emission
lines, and to better distinguish random clumping of photon noise 
from weak absorption.

The size of the error associated with the DA at a given $z$ also
depends on the rate of change of DA with $z$. 
If DA(z) were a power law over a wide range of $z$, and our
measurements were unbiased at all $z$, then the error obtained using
the power law fit will be very small because we can use points from
all $z$ together.  
Alternatively, if the true DA(z) departs from a power law, or if the
DA that we measure at some $z$ is biased, then the error could be
systematically larger than we might expect from the plots, tabulated values
and errors.

In \citet{tytler04b} we showed that most of the scatter in the DA at a 
given $z \sim 2$ was due to large scale structure.
However, some of the scatter in Figs. \ref{fosonlyda}
and \ref{allda} could come from variable amounts of systematic error,
such as the continuum level too high in one QSO and too low in another, or
objects in which we failed to identify and mask metal lines. The
scatter in the points at a given $z$ appears non-Gaussian, with an
excess of large deviations. We would not be surprised if some of the
outliers had relatively low SNR, leading to unusual continuum level errors, 
and in the case of the high points, unidentified metal lines.

Overall, the metal lines are probably a smaller source of error in
the IGM DA than is the continuum level. Even if the true DA from metal
lines were twice our estimates, this would reduce the DA values for
the IGM in Table \ref{tablefosda} by a factor of 0.85.  The STIS
spectra also indicate that missing metal lines are not the major
error. We obtained similar DA values from FOS and STIS spectra of the
same QSOs (Fig. \ref{stisda}), where we conducted an unusually
thorough search for metals in the STIS spectra \citep{milutinovic06a}.

In this paper we also measured the flux autocorrelation and the PDF of
the flux in the FOS spectra. Both depend on the amount of absorption
(DA) as well as the factors that control the distribution of the
absorption in wavelength or velocity.  These factors include the gas
temperature, the Hubble constant, large scale structure and the
SNR and spectral resolution. For the FOS spectra the spectral resolution is
the dominant factor in the shape of the flux autocorrelation and
PDF. However, the FOS sample is large enough that we can measure
non-zero autocorrelation out to much larger velocity lags than have
been seen in STIS spectra.

The most obvious ways to improve the measurements of the absorption by
H~I in the IGM would be to include high SNR spectra of tens of QSOs at 
$z \simeq 1.6 - 1.8$ with complete wavelength coverage, and to have 
automated continuum fitting methods that are more accurate than manual methods. 

\section*{Acknowledgments}

This work made extensive use of the FOS spectra and line
identifications prepared by Jill Bechtold, Jennifer Scott, and Adam
Dobrzycki (http://lithops.as.arizona.edu/$\sim$jill/QuasarSpectra/).  This
work was supported by HST-AR-10288.01. The NSF REU program supported 
Samuel Bockenhauer who helped identify metal lines in the STIS spectra.

\bibliographystyle{mn2e}
\bibliography{archive}

\clearpage

\end{document}